\documentstyle[aps,epsf,prd,epsfig]{revtex}

  \newcommand{\be}{\begin{equation}}
  \newcommand{\ee}{\end{equation}}
  \newcommand{\bea}{\begin{eqnarray}}
  
  \newcommand{\eea}{\end{eqnarray}}
  \newcommand{\nn}{\nonumber}
  \newcommand{\dd}{\displaystyle}
  \newcommand{\spur}[1]{\not\! #1 \,}
\begin{document}
 \title{
  \hfill{\small \begin{tabular}{l}
  BA-TH/404-00\\
  Napoli-DSF~2000/37\\
   \\ \\ \\
   \end{tabular}}\\
{\LARGE\bf Charming penguin contributions to $B\to K \pi $ } }
\author{{\bf C. Isola$^{a}$, M. Ladisa$^{a,b}$, G. Nardulli$^{b}$, T. N. Pham$^{a}$,
P. Santorelli$^{c}$}
}
\maketitle
\begin{it}
\begin{center}
   $^a$Centre de Physique Th{\'e}orique, \\
  Centre National de la Recherche Scientifique, UMR 7644, \\ {\'E}cole
  Polytechnique, 91128 Palaiseau Cedex, France\\  \vspace*{0.3cm}
  $^b$Dipartimento di Fisica dell'Universit{\`a} di Bari, Italy\\
  Istituto Nazionale di Fisica Nucleare, Sezione di Bari,
  Italy\\\vspace*{0.3cm}
  $^c$Dipartimento di Scienze Fisiche, Universit{\`a} di
  Napoli "Federico II", Italy\\Istituto Nazionale di Fisica Nucleare, Sezione di
  Napoli, Italy\vspace*{0.3cm}
\end{center}
\end{it}
\begin{abstract}
We present calculations of the charming-penguin long-distance
contributions to $B \to K\pi$ decays due to intermediate  charmed
meson states. Our calculation is based on the Chiral Effective
Lagrangean for light and heavy mesons, corrected for the hard pion
and kaon momenta. We find that the charming-penguin contributions
increase significantly the $B\to K\pi$ decay rates in comparison
with the short-distance contributions, giving results in better
agreement with experimental data.
\end{abstract}
\thispagestyle{empty}
\setcounter{page}{1}
\newpage
\section{Introduction}
 \par
 \noindent
 Two-body
charmless $B$ decays are studied as a mean to detect direct
CP-violation and to determine the CKM mixing parameters in the
Standard Model \cite{babarbook}. In these processes the short-distance
contributions can be computed by using the factorization
approximation for the local operators in the effective
non-leptonic  Hamiltonian ${\cal
  H}_{\rm eff}$ as  argued by Bjorken on the basis of color-transparency
 \cite{Bjorken} and  recently justified in \cite{Beneke}.
However non-leading ${\cal O}(1/m_b)$ effects which appears in the
penguin matrix elements may be numerically important. In some
cases where experimental data are available this approach does
meet difficulties (for a discussion see e.g. \cite{Ciuchini1}). An
example is offered by the decay $B\to K\pi$. The branching ratios
for these processes have been measured by the CLEO collaboration
\cite{cleo}:
\begin{eqnarray}
{\mathcal B} (B^{+} \rightarrow K^0 \pi^{+})  &=&
(18.2^{+4.6}_{-4.0}\pm 1.6) \times 10^{-6},   \nonumber  \\
{\mathcal B} (B^0\rightarrow K^{+} \pi^{-})  &=&
(17.2^{+2.5}_{-2.4}\pm 1.2) \times 10^{-6}, \label{CLEO2}
\end{eqnarray}
and, more recently, by
the  BaBar \cite{Babar} and Belle \cite{Belle} collaborations.
BaBar gives a value of ${\mathcal B} (B^0\rightarrow K^{+}
\pi^{-}) = (12.5^{+3.0+1.3}_{-2.4-1.7}) \times 10^{-6}$. Belle
gives\footnote{After having completed this work new data from
the BaBar \cite{cavoto} and Belle \cite{casey} Collaborations have
been published: BaBar gives ${\mathcal B} (B^{+} \rightarrow K^0 \pi^{+})\ =\
(18.2^{+3.3+1.6}_{-3.0-2.0}) \times 10^{-6}$ and
${\mathcal B} (B^{0} \rightarrow K^+ \pi^{-})\ =\
(16.7\pm 1.6^{+1.2}_{-1.7}) \times 10^{-6}$. Belle gives
${\mathcal B} (B^{+} \rightarrow K^0 \pi^{+})\ =\
(1.31^{+0.55}_{-0.46}\pm 0.26) \times 10^{-5}$ and
${\mathcal B} (B^{0} \rightarrow K^+ \pi^{-})\ =\
(1.87^{+0.33}_{-0.30}\pm 0.16) \times 10^{-5}$.\
Within the errors they are compatible with previously released data.}
${\mathcal B} (B^0\rightarrow K^{+} \pi^{-}) =
(17.4^{+5.1}_{-4.6}\pm 3.4) \times 10^{-6}$ and ${\mathcal B}
(B^{+} \rightarrow K^0 \pi^{+}) = (16.6^{+9.8 +2.2}_{-7.8-2.4} )
\times 10^{-6} $. If one now evaluates  the amplitudes for these
decays by including only  tree and penguin operators, without
taking into account the charm quark loop, the factorization
approximation produces  branching ratios too small as compared to
the data \cite{Ciuchini1}.

To get a better agreement one is forced to include the so-called
charming penguin operators, i.e. those operators which, being
proportional to the large CKM factor $V_{cb}^* V_{cs}$ and the
large Wilson coefficient $c_2\approx 1$, are not suppressed like
the tree and penguin contributions. Two approaches can be followed
to include these effects. One may define effective Wilson
coefficients by considering the effect of a charm quark loop
treated perturbatively \cite{gerard,Fleischer,Fleischermannel,Deshpande1,Ali};
in this way also
an absorptive part of the non-leptonic decay amplitude is
generated. This approach, used together with the factorization
approximation, seems to produce decay rates in agreement with
data, at least qualitatively, as shown in some previous papers
\cite{Ali,Deandrea,Deshpande,Isola}.
In these works the inclusion of the charm
quark loop increases the effective Wilson coefficients of the
strong penguin operators by about $30\%$, thereby producing $B \to
K\pi$ decay rates closer to the data. More recently charm quark
effects computed by this method have been included in works
dealing with the validity of the factorization
\cite{Beneke,Du,Muta}.

A different approach can however be followed. It assumes that the
charm quark contributions are basically long-distance effects that
can be taken into account by including rescattering processes such
as, e.g. $B\to D D_s\to K\pi$. These contributions, first
discussed to our knowledge  in \cite{Nardulli}, have been more
recently stressed by \cite{Ciuchini1}, where they are called
charming penguin terms, a nomenclature we shall adopt here.

The aim of this paper is to present an evaluation of the charming
penguins in the $B\to K\pi$ decays going beyond the
parametrizations of \cite{Ciuchini1}.
As a matter of fact, instead of considering only $D,D_s$ intermediate
states, we will also consider charmed vector mesons \footnote{The
transition $B \to X_c$ is saturated by $X_c~=~D + D^*$ at zero
recoil \cite{SV}. Though we are not in this limit, the
approximation can be used as a guideline.}. Furthermore, by
using phenomenological information from semileptonic decays and
Chiral Effective Lagrangean,  we shall estimate both the real and
the imaginary part of the charming penguins. This result, being an
improved determination of the strong phase, might be of some utility
in connections with strategies to determine the angle $\gamma$ of the
unitarity triangle from future more precise data.

The long-distance absorptive part  is essentially due to the
rescattering effects of the processes $D^{(*)} D^{(*)}_{s} \to
K\pi$. The situation is similar to the $B_{s} \to \gamma\gamma$
decay for which  the absorptive part obtained in \cite{Choudhury}
is comparable to the short-distance contribution. As noted above,
the dispersive part of the charming penguins, in the present
approximation, was computed previously in \cite{Nardulli} for a
number of charmless $B$ decays to two pseudo-scalar mesons as well
as one pseudo-scalar and one vector meson in the final states. In
this paper we present a new calculation of the long-distance
contribution using the same Cottingham formula \cite{Cottingham},
and more recent information on the semileptonic decay form
factors.

The plan of the paper is as follows. In the next section we
discuss the separation between the short-distance and the
long-distance contributions. In section 3, the absorptive part is
shown in terms of the $B \to D^{(*)} D_s^{(*)} $ and $  D^{(*)}
D_s^{(*)}\to K\pi $ amplitudes obtained by the short-distance
non-leptonic Hamiltonian and the $B \to D^{(*)}$ and $D^{(*)} \to
K\pi$ semileptonic decay form factors. Section 4 is  devoted to
the calculation of the dispersive part (the real part) of the $B
\to K\pi$ amplitudes. Finally in section 5 we compare our results
to the data and draw our conclusions.

\section{Short-Distance and Long-distance non-leptonic weak matrix elements }
 The non-leptonic $B \to K\pi$ decay
amplitude is obtained by considering  the matrix element
\begin{equation}
A_{K \pi}~=~<K(p_{K})\pi(p_{\pi})|i{\cal H}_{\rm eff}|B(p_{B})> \; .
\end{equation}
The effective Hamiltonian for non-leptonic $B$ decays is the sum
of 4-quark tree-level and penguin operators and is given by
\begin{equation}
{\cal H}_{\rm eff} = {G_{F} \over \sqrt{2}} \left[V_{ub}^*
V_{us}(c_1 O_1^u + c_2 O_2^u) + V_{cb}^* V_{cs}(c_1 O_1^c + c_2
O_2^c) \nonumber \\
- V_{tb}^* V_{ts}\left( \sum_{i=3}^{10} c_i O_i + c_g O_g
\right)\right]
\label{Heff}
\end{equation}
where $c_i$ are the Wilson coefficients evaluated at the normalization
scale $\mu = m_b$ \cite{Fleischer,Buras,Ciuchini} and
next-to-leading QCD radiative
corrections are included. $O_1$ and $O_2$ are the usual tree-level
operators, $O_i$ ($i=3,..., 10$) are the penguin operators and
$O_g$ is the chromomagnetic gluon operator. The ${\it c_i}$ in eq.
(\ref{Heff}) are as follows \cite{Buras}:
$c_2=1.105,~c_1=-0.228,~c_3=0.013,~c_4=-0.029,~c_5=0.009,
~c_6=-0.033,~c_7/\alpha=0.005,~c_8/\alpha=0.060,
~c_9/\alpha=-1.283,~c_{10}/\alpha=0.266$.

In the calculation of the $B \to K\pi$ decay amplitude $A_{K\pi}$
we can separate the short-distance and the long-distance
contributions:
\begin{equation}
A_{K\pi} = A_{SD} + A_{LD} \label{ampl}\ .
\end{equation}
The short-distance part of the amplitude  $A_{SD}$ arises from the
operators in (\ref{Heff}) that give non vanishing contributions in
the factorization approximation, i.e. $O_i^u$ ($i=1,2$) and $O_i$
($i=3,...,10$). In this approximation  it is given by:
\begin{eqnarray}
A_{SD}(B^+ \to K^0 \pi^+) &=& \frac{G_F}{\sqrt{2}}\ f_K\ F_0^{B
\to \pi}(m_K^2)\ (m_B^2-m_\pi^2) \times \nn \\ & \times & V_{tb}^*
V_{ts} \left[\
   a_4 - \frac{1}{2} a_{10}  +\left(\ a_6 - \frac{1}{2} a_{8} \right )
  \frac{2\ m_K^2\ }{(m_b - m_d)\ (m_d
      + m_s)}\  \right]\ ,  \\
A_{SD}(B^0 \to K^+ \pi^-) &=& -\frac{G_F}{\sqrt{2}}\ f_K\ F_0^{B
\to \pi}(m_K^2)\ (m_B^2-m_\pi^2) \times \nn \\ &\times& \left[\
V_{ub}^* V_{us}\ a_2 -  V_{tb}^* V_{ts}\
  \left(\ a_4 + a_{10} + \frac{2\ m_K^2\ (a_6 + a_8)}{(m_b - m_u)\ (m_u
      + m_s)}\ \right)\ \right]  ,
\end{eqnarray}
where ${\dd a_i = c_i + \frac{c_{i+1}}{3}}$ (i=odd) and ${\dd a_i
= c_i + \frac{c_{i-1}}{3}}$ (i=even).

\begin{figure}[ht!]
\begin{center}
\epsfig{file=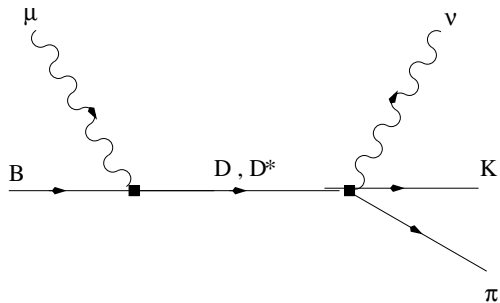,height=4cm}
\end{center}
\caption{{\small The diagram corresponding to the hadronic tensor
$T_{\mu\nu}$  in equation (\ref{Tmn}).
The boxes represent weak couplings.}}
\label{B-Kpi-diagram}
\end{figure}

Numerically, for $|V_{ub}|=~0.0038,~V_{us}=~0.22,~V_{tb}\simeq
1,~V_{ts}=-\,0.040$ and $\gamma$ = $-\,arg \left(V_{ub}\right)=$
$~54.8^o$\cite{ciuchini3} and $F_0^{B\to \pi}(m_K^2)\ =\ 0.37$,
we get
\begin{eqnarray}
A_{SD}(B^+ \to K^0 \pi^+) &=& 2.43\times 10^{-8}~{\rm GeV}\cr
A_{SD}(B^0 \to K^+ \pi^-)&=& \left( 1.86 -i\, 0.95\right) \times
10^{-8}~{\rm GeV}~.
\end{eqnarray}
As discussed in the introduction by this contribution alone we
would obtain  branching ratios roughly one order of magnitude
smaller than the experimental findings. Therefore one has to relax
some of the hypotheses. The  possibility we explore here  is to
consider non-factorizable contributions. In this case
 also the operators $O_i^c$  ($i=1,2$)
are effective; these terms will be treated as long-distance
contributions, i.e. we will go beyond vacuum saturation and
consider intermediate low energy hadronic states in the product of
the weak currents in the operators of (\ref{Heff}).  Also other
operators in (\ref{Heff}) have long-distance contributions, but
clearly $A_{LD}$ is dominated by $O_2^c$  due to the enhancement
of the CKM factor $V_{cb}^* V_{cs}$ and the Wilson coefficient
$a_2$. Therefore we can write:
\begin{eqnarray}
A_{LD} &=& A_{LD}(B^+ \to K^0 \pi ^+) ~=~ A_{LD}(B^0 \to K^+ \pi
^-)~=~ \nn \\ &=& {G_F \over \sqrt{2}}\,V_{cb}^* V_{cs}\, a_2\int
\,{d^4q \over (2\pi)^4}\theta(q^2 + \mu^2)\,T(q,p_B,p_K,p_{\pi})
\label{ald}
\end{eqnarray}
where $\mu$ is a cut-off separating long-distance and
short-distance contributions, while the amplitude
$T(q,p_B,p_K,p_{\pi})= g^{\mu\nu}\,T_{\mu\nu}$ is obtained by:
\begin{equation}
T_{\mu\nu} ~=~i\,\int \,d^{4}x\,\exp (i\, q\cdot x)
<K(p_K)\pi(p_\pi)|{\rm T} (J_{\mu}(x)J_{\nu}(0))|B(p_B)>
\label{Tmn} ~,
\end{equation}
with $J_{\mu} = \bar{b}\gamma_{\mu}(1 - \gamma
_5)c$ and $J_{\nu} = \bar{c}\gamma_{\nu}(1 - \gamma _5)s$.
$A_{LD}$ has the same value for the two channels $B^+ \to K^0 \pi
^+,~ B^0 \to K^+ \pi ^-$. We will saturate the $T-$product of the
two currents by inserting $D,~D^*$ intermediate states, see fig.
\ref{B-Kpi-diagram}.

\vspace{1truecm}

\begin{figure}[ht]
\begin{center}
\epsfig{file=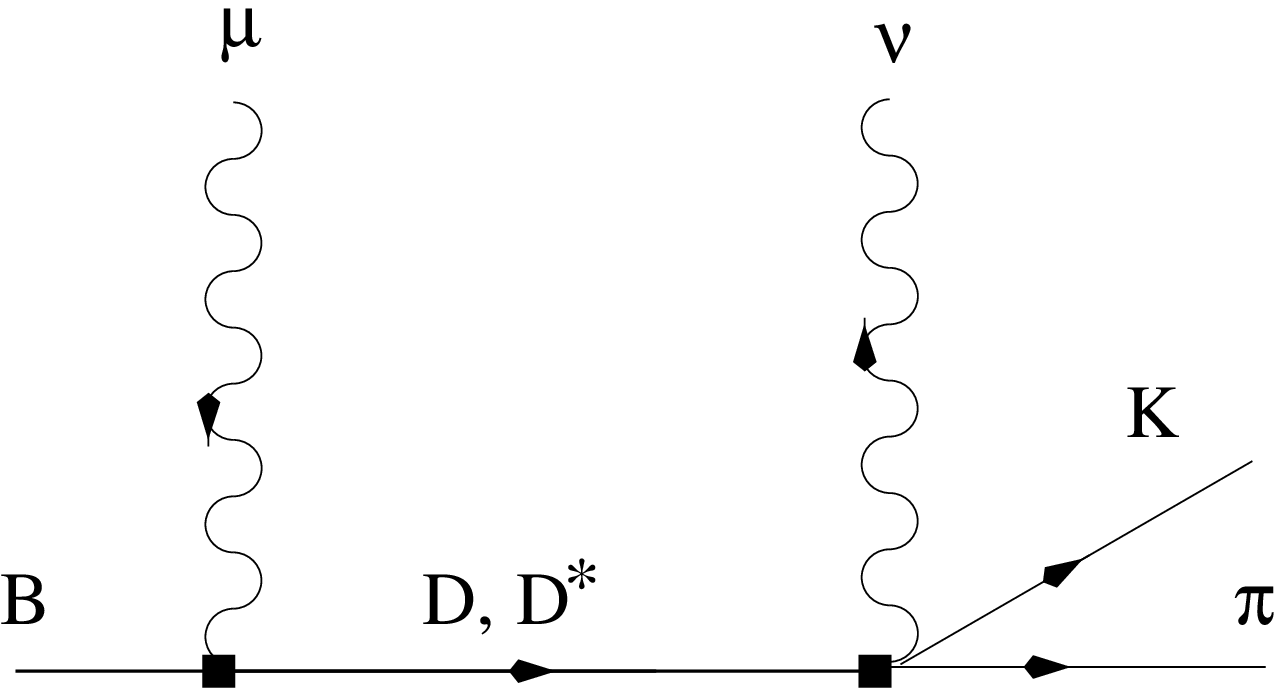,height=3.5cm} \hspace*{2cm} \epsfig{file=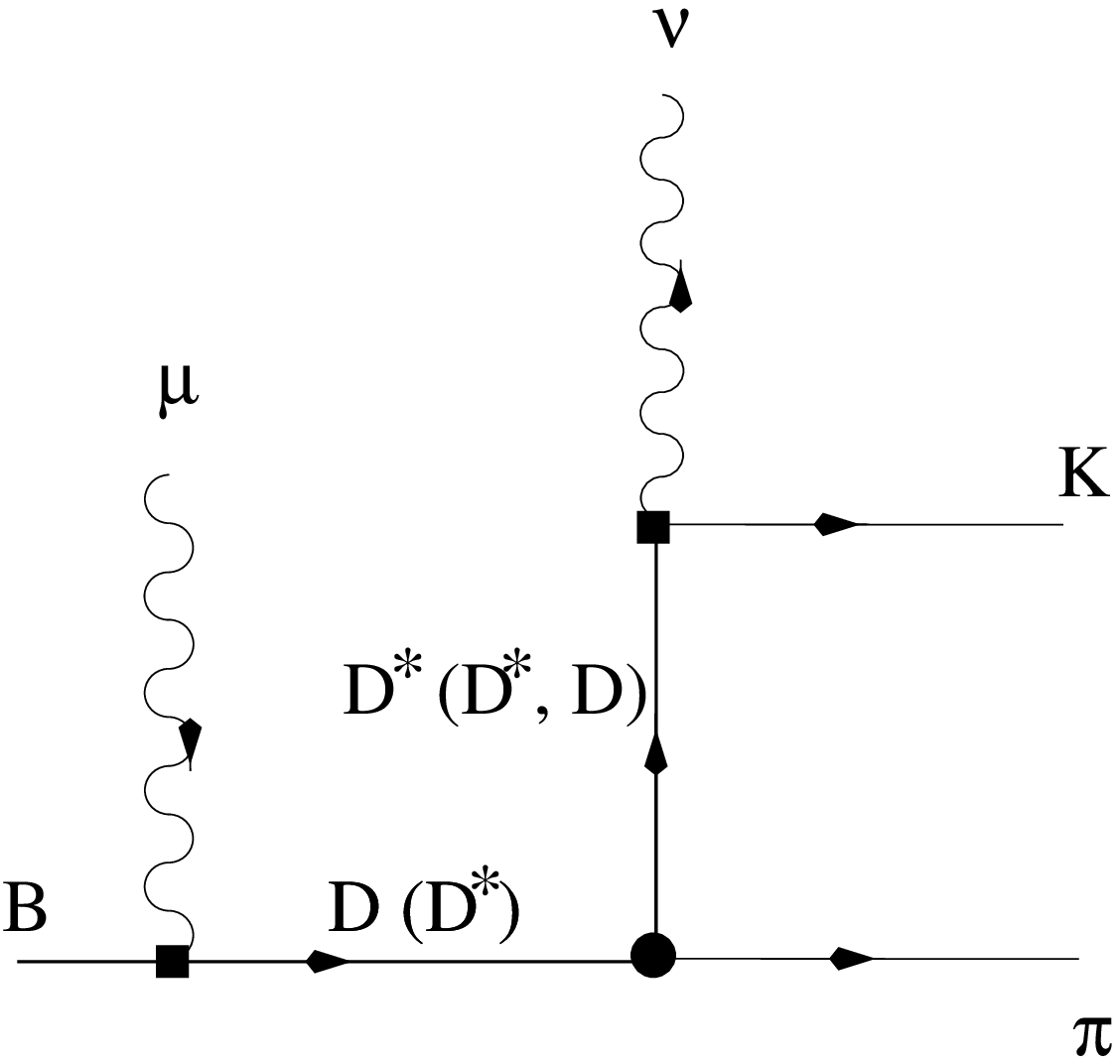,height=5cm}\\
(a) \hspace*{6.5cm} (b) \\
\epsfig{file=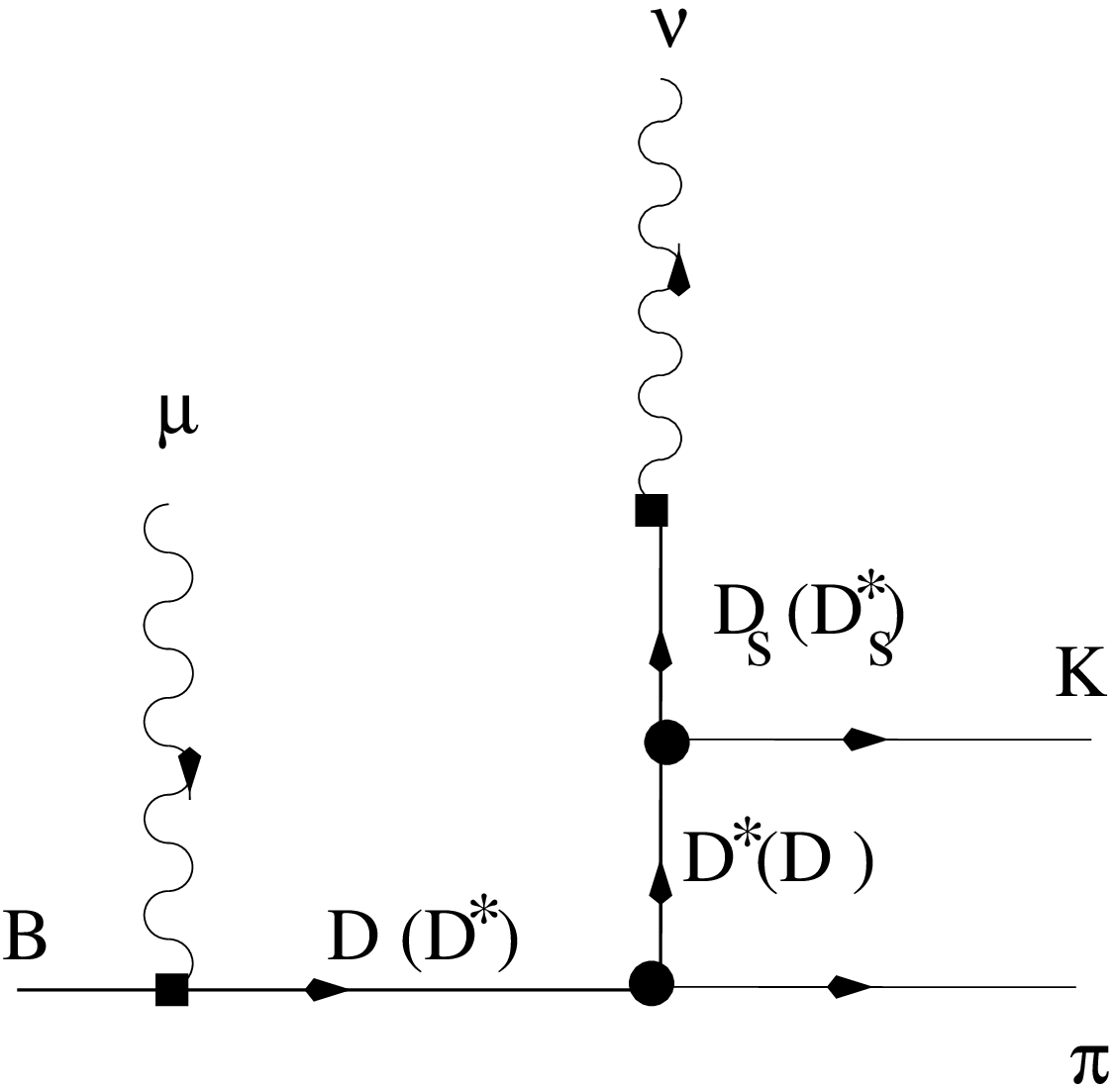,height=4.5cm} \hspace*{2cm} \epsfig{file=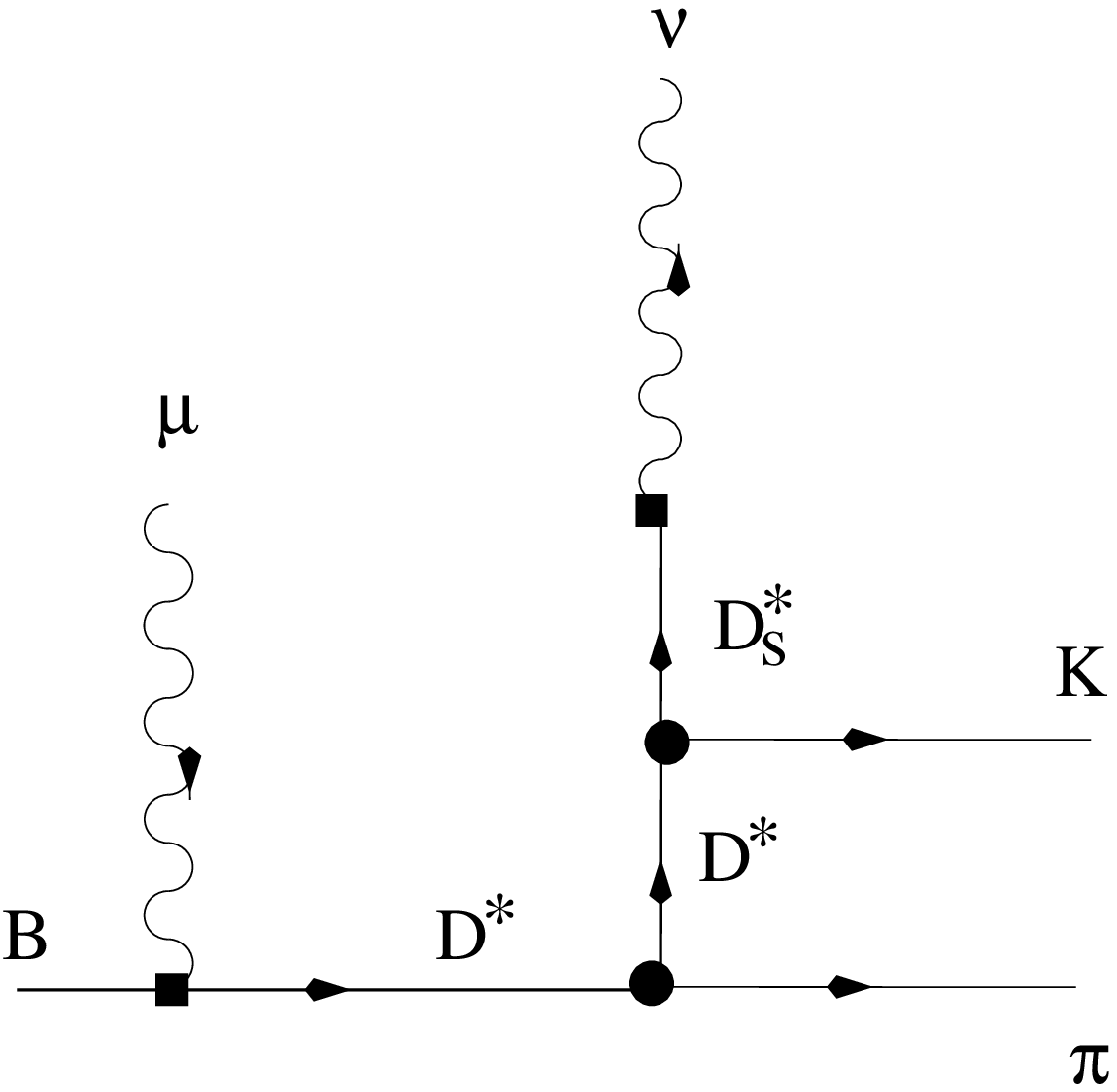,height=4.5cm}\\
(c) \hspace*{6.5cm} (d) \\
\epsfig{file=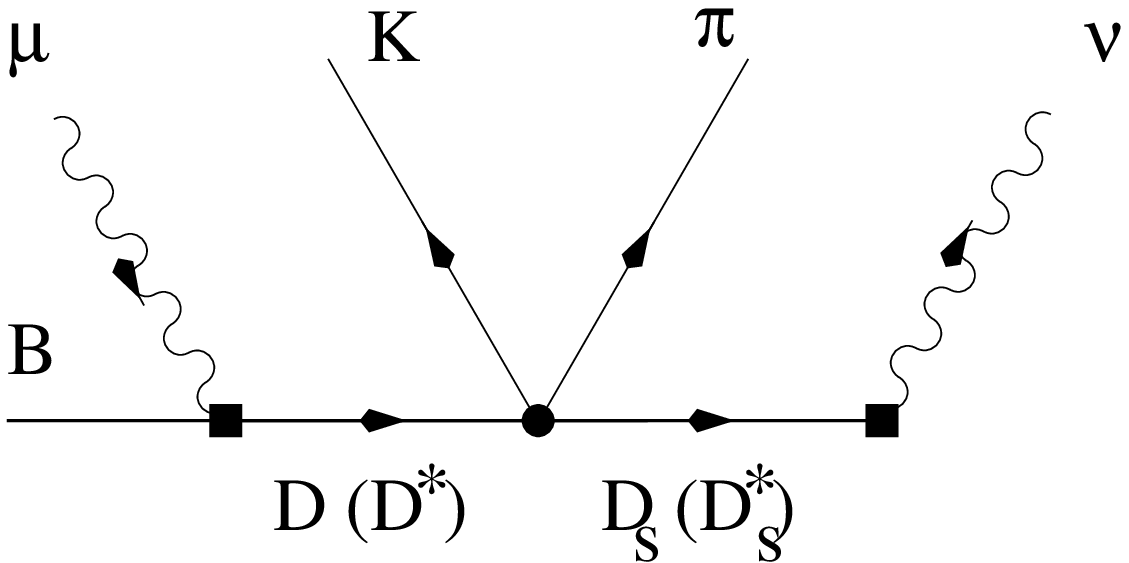,height=3cm} \hspace*{1cm} \epsfig{file=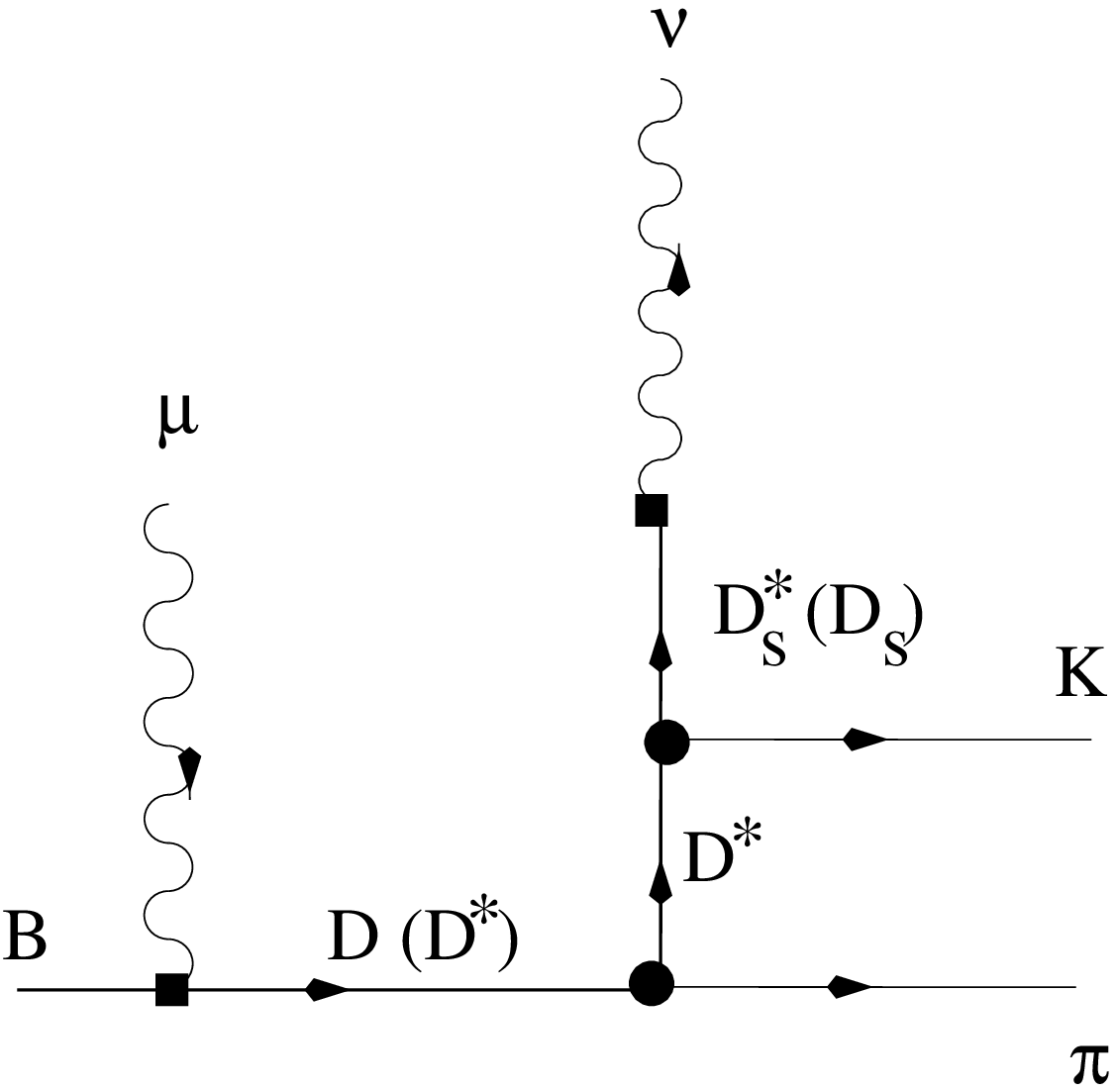,height=4.5cm}\\
(e) \hspace*{6.5cm} (f)
\end{center}
\caption{{\small The relevant Feynman diagrams used to compute
$T_{\mu \nu}$. The box and the circle refer, respectively, to
  the weak and to the strong interaction. Fig. (b) corresponds to three
  diagrams: the first one has the $D$ state on the horizontal
  internal line and a $D^*$ state on the vertical line; the second one
  presents a $D^*$ state on both lines; the third graph has, respectively, a $D^*$ and a
  $D$ state. The same rule applies
  to the other figures.}} \label{ff2}
\end{figure}

\begin{figure}[h]
\begin{center}
\epsfig{file=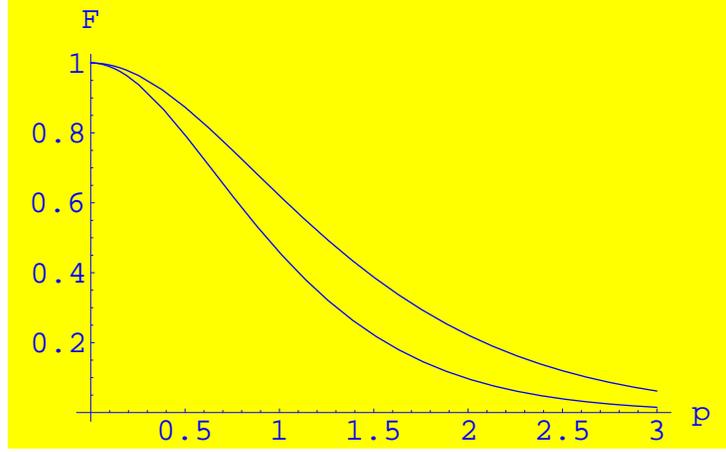,height=6cm}
\end{center}
\caption{{\small Form factor $F(p)$ for the $D^*D\pi$ strong
coupling as a function of the pion momentum $p=|\vec p_\pi|$ in
GeV. Normalization is for the soft pion limit. The two curves
correspond to $<k>=400$ MeV (upper) and $<k>=300$ MeV (lower).}}
\label{ff}
\end{figure}

To compute the diagram of fig. \ref{B-Kpi-diagram} one has to
model the $B\to D^{(*)}$ and the $D^{(*)}\to K\pi$ weak
transitions; this will be done following the  Chiral Effective
Lagrangean approach of \cite{lee}. We list here the effective
vertices and currents of this theory.
\begin{itemize}
\item[1)]
Strong coupling among two heavy and one light meson:
\begin{equation}
{\cal L}_{HH \pi}~=~\frac{i g}{2}\ Tr \overline{H}_a H_b
 \gamma ^\mu\ \gamma _5 \left( \xi ^\dag \partial _\mu \xi - \xi
 \partial _\mu \xi ^\dag \right)_{ba} \; ,
\label{strong-3}
\end{equation}
where
\begin{equation}
H_a ~=~ \frac{1+\spur{v}}{2}\ \left( P^*_{a \mu} \gamma ^\mu - P_a
  \gamma _5 \right) \;\; ,~~~~~~~~\overline{H}_a ~=~ \gamma ^0 H^\dag_a
  \gamma ^0 \;\; ,
\end{equation}
and
\begin{equation}
\xi ~=~ exp\left\{i \frac{{\cal M}}{f}\right\} \;\; .
\end{equation}
Here $v$ is the heavy meson velocity, $P_a,P^*_{a \mu}$ are the
annihilation operators of heavy pseudo-scalar and vector mesons
made up by a heavy quark and a light antiquark of flavour $a$
($a~=~1,2,3$ for $u,~d,~s$); ${\cal M}$ is the usual $3 \times 3$
matrix comprising the octet of pseudo--Goldstone bosons; $f$ is
the pseudo--Goldstone bosons decay constant ($f \approx
f_\pi\approx 130$ MeV).
\item[2)]
Strong coupling among two heavy and two light mesons:
\begin{equation}
{\cal L}_{HH \pi \pi}~=~\frac{i}{2}\ Tr \overline{H}_a H_b
 v ^\mu\ \left( \xi ^\dag \partial _\mu \xi + \xi
 \partial _\mu \xi ^\dag \right)_{ba} \; .
\label{strong-4}
\end{equation}
\item[3)]
Weak coupling of a heavy meson to pseudo--Goldstone bosons by a
${\bar q}_a\gamma^\mu(1-\gamma_5)Q$ current:
\begin{equation}
L_{\mu a}~=~\frac{i \alpha}{2}\ Tr \gamma _\mu (\ 1 - \gamma _5\ )
H_b
 \xi ^\dag_{ba} \; .
\label{weak-current}
\end{equation}
$\alpha$ is related to the heavy meson leptonic decay constant
by the formula $\alpha=$ $f_H\ \sqrt{m_H}$, valid in the infinite
quark mass limit. Eq.(\ref{weak-current}) generates, for example,
weak couplings of $D,~D^*$ to hadronic final states with
$0,1,2,...$ pseudo--Goldstone bosons.
\item[4)]
The weak matrix elements $< ( D ,\ D^* ) |J^\mu | B >$ are
parameterized as in \cite{lee}, i.e. in the infinite heavy quark
limit and introducing the Isgur--Wise function for which we use
the simple expression
\begin{equation}
\xi_{IW}(v \cdot v^\prime) ~=~ \xi_{IW}(\omega) =
\left(\frac{2}{1+\omega}\right)^2\, , \label{IW}
\end{equation}
where $v,v^\prime$ are heavy meson velocities.
\item[5)]
The weak matrix elements $< K \pi | J^\mu | D>$ and  $< K \pi |
J^\mu | D^*>$ can be computed by the rules given above, i.e. using
the model of \cite{lee}. It amounts  to consider polar diagrams as
well as direct production of light mesons in the framework of the
Chiral Effective Lagrangean \footnote{We do not include graphs
where the pion is emitted at the first vertex; either they are
already taken into account by the contributions discussed so far,
and one must omit them to avoid double counting, or they are
negligibly small, according to the estimates in \cite{Nardulli}.}.
This corresponds to the evaluation of the diagrams in fig.
\ref{ff2}.
\end{itemize}
\newpage
Strictly speaking, the evaluation of these diagrams by the Chiral
Effective Lagrangean is valid only for soft light pseudo-scalar
mesons. Therefore, in order to use it in the present context, we
have to account for the high momenta of the outgoing light mesons.
To this aim we introduce two modifications: {\bf i)} we keep the
full propagator in the pole contributions instead of the
expressions in the soft pion limit (a similar use of the full
$D^*$ propagator to go beyond the soft pion result
has also been given in \cite{Fajfer}) ; {\bf ii)} we introduce a form
factor in the strong coupling constant of light and heavy mesons
(a similar approach is used in semileptonic decays
\cite{casalbuoni}). Let us consider the $D^*D\pi$ coupling that
can be written in general as follows:
\begin{equation}
\langle D(p^\prime)\,
\pi(p_\pi)|D^*(p,\epsilon)\rangle~=~G_{D^*D\pi}\ (\epsilon\cdot
p_\pi)~.
\label{G}
\end{equation}
From (\ref{strong-3}) we have
\begin{equation}
G_{D^*D\pi}=\frac{2m_D g}{f_\pi}~,
\label{GG}
\end{equation}
where $g$ is independent of the heavy flavor and is predicted to have the value
(see \cite{casalbuoni} and references therein):
\begin{equation}
g\approx 0.40~.
\end{equation}
We shall neglect $1/m_Q$ effects in this case and shall
adopt this value. There is, however, an important point to be
discussed here. The expression (\ref{G}) is derived from the low
momentum  Chiral Effective Lagrangean (\ref{strong-3}). The
neglect of higher pseudo-Goldstone boson derivatives is however
not justified in our case, as the pion and kaon have momenta of
the order of $m_B/2$. To include this effect we modify (\ref{GG})
as follows:
\begin{equation}
G_{D^*D\pi}=\frac{2m_D g}{f_\pi}F(|\vec p_\pi|)~,
\label{GGG}
\end{equation}
where $F(|\vec p_\pi|)$ is a form factor
normalized as $F(0)=1$. This form factor can be evaluated by using
the constituent quark model. For this purpose, let us introduce
the heavy meson wavefunction in the momentum space: $\psi_D(k)$,
where $k=|\vec k |~=~|\vec q_1-\vec q_2|/2$ is one half of the
relative momentum of the two component quarks (whose momenta are
respectively $\vec q_1$ and $\vec q_2$). The coupling constant $g$
in the soft pion limit is proportional to the overlap of the $D^*$
and $D$ wave functions:
\begin{equation}
g~\propto \int d^3k\ \psi^*_D (|\vec
k|) \psi_{D^*}(|\vec k|)f(k)
\end{equation}
where $f(k)$ is some smooth
function whose precise shape depends on the particular model one
employs\footnote{For a particular calculation using this approach
see \cite{defazio}.}, but it is irrelevant for our purposes. We
therefore get, if the pion momentum is $\vec p_\pi$
\begin{equation}
F(|\vec p_\pi|)~=~ \frac{ {\dd\int d^3k\ \psi^*_D \left( \Big |\vec
k-\frac{\vec p_\pi}2 \Big| \right) \psi_{D^*}(|\vec k|)f(k)}}{\dd
{ \int d^3k\ \psi^*_D (|\vec k|) \psi_{D^*}(|\vec k|)f(k)   }} ~,
\label{GGGG}
\end{equation}
where the denominator has been introduced to normalize correctly
the form factor. We employ for the wave function the
expression\footnote{This expression is sufficiently general; it
also corresponds to  fits of the wave function in particular
constituent quark models, see e.g. \cite{Brho,Bpi}.}:
\begin{equation}
\psi^*_D (k)=\psi^*_{D^*} (k)\propto e^{-\frac{\alpha k}2}
\end{equation}
which corresponds to an average quark momentum
\begin{equation}
< k> =\frac{3}{\alpha}
\end{equation}
inside the meson and we assume a constant
value for the smooth function $f(k)$.  We plot in fig. \ref{ff}
the value of the form factor as a function of the pion
three-momentum for two values of $\alpha$, corresponding,
respectively to $< k >=400~{\rm MeV}$ (upper curve) and  $< k
>=300~{\rm MeV}$ (lower curve).
For $|\vec
 p_\pi|\simeq m_B/2$ we get therefore
\begin{equation}
 F(|\vec p_\pi|)~=~0.065\pm 0.035~,
\label{fpi}
\end{equation}
where the central value corresponds to $<k>=350$ MeV
and the higher (lower) to $<k>=400$ ($<k>=300$) MeV.

It should be observed that also the Lagrangean ${\cal L}_{HH \pi
\pi}$ and the weak current $L^\mu$ have corrections from terms
containing higher order derivatives. However these terms do not
contribute to the imaginary part of $A_{LD}$; on the other hand
hard pion and kaon effects in the calculation of to the real part
can be taken into account by an appropriate cut-off of the
Cottingham formula. We will discuss the problem in section 4.

\section{Imaginary part of the long-distance contribution}
\par\noindent
It can be easily seen that only diagrams \ref{ff2}c and \ref{ff2}d
contribute to the discontinuity of $A_{LD}$. The diagrams
\ref{ff2}a and \ref{ff2}b have no discontinuity, whereas the
diagram \ref{ff2}e vanishes in the chiral limit and we neglect it
altogether. As to the diagram \ref{ff2}f, both its imaginary and
its real part vanish, as  it can be easily checked.

We use the following kinematics:
\begin{equation}
p^\mu~=~m_B v^\mu=(m_B,\vec 0
)\ ,\ ~~~~ p^\mu_{D^{(*)}}~=~m_{D}v^{\prime\mu}\ ,\ ~~~~
q~=~p-p_{D^{(*)}}\ .
\end{equation}
The discontinuity of the diagrams of
figs. (\ref{ff2}b, \ref{ff2}c) gives
\begin{eqnarray}
{\rm Disc}\ A_{LD} &=&
 2\, i\, {\rm Im}\, A_{LD}=\, (-2\pi i)^2 \int\frac{d^4
q}{(2\pi)^4}\delta_+(q^2-m^2_{D_s})\, \delta_+(p^2_{D^{(*)}}-m^2_D
)\times \cr &&\cr&\times & A(B\to D_s^{(*)}D^{(*)})\,
A(D_s^{(*)}D^{(*)}\to K\pi)\ =\cr&&\cr &=& \, -\frac{m_D}{16\pi^2
m_B}\, {\sqrt{\omega^{*2}-1}}\, \int d\vec n \, A(B\to
D_s^{(*)}D^{(*)})\, A(D_s^{(*)}D^{(*)}\to K\pi)\ ,
\end{eqnarray}
where
$\dd
\omega^* \ =\ \frac{m_B^2+m_D^2-m^2_{D_s}}{2 m_D m_B}$ and the
angular integration is over the directions of the vector $\vec
v^{\prime}=\vec n \sqrt{\omega^{^* 2}-1}$. Our results are as
follows:
\begin{eqnarray} A(B\to D_s D)&=& -\, K\, (m_B-m_D)(1+\omega^*)\ ,
\cr \cr
 A(B\to D_s^{*}( \eta, \, q )D^{*}( \epsilon, v^\prime ))&=& K\,
m_{D_s}\eta^{*\mu}\epsilon^{*\alpha}\left(i\epsilon_{\alpha\lambda\mu\sigma}
v^{\prime\lambda}v^\sigma-g_{\mu\alpha}(1+\omega^*)+v_\alpha
v_{\mu}^\prime \right)\ ,
\end{eqnarray}
where $\dd K\ =\ \frac{G_{F}}{\sqrt
2}V_{cb}^* V_{cs}\ a_2 \sqrt{m_B m_D} f_{D_s}\xi_{IW}(\omega^*)\
.$
On the other hand we have
\begin{eqnarray}
A(D_s D\to K\pi)&=&\ i\
\frac{\left(2 g\,m_{D}\, F(|\vec p_\pi|)\right)^2 }{f_\pi f_K}\
\times \cr&&\cr &&\frac{-p_\pi\cdot p_K+p_\pi\cdot(q-p_K)(p_K\cdot
q-m^2_K)/m^2_{D^*}}{ (m_D v^\prime-p_\pi)^2-m^2_{D^*} }\ , \cr &&
\cr A(D_s^{*}(\eta ,q) D^{*}(\epsilon, v^\prime) \to K\pi)&=& \ i\
\frac{\left(2 g\, F(|\vec p_\pi|)\right)^2 {\sqrt{m_{D_s^*}
m_{D^*} } } }{f_\pi f_K}\ \epsilon_\lambda\eta_\sigma \ \times
\cr&&\cr &&\left( \frac{m_{D^*}\ p_\pi^\lambda\, p_K^\sigma}{
(m_{D^*}v^\prime-p_\pi)^2-m^2_{D} }\ +\
\frac{G^{\sigma\lambda}(p_\pi,p_K, v^\prime)}{
(m_{D^*}v^\prime-p_\pi)^2-m^2_{D^*} }\right),
\end{eqnarray}
where
\begin{eqnarray} G^{\sigma\lambda}(p_\pi,p_K, v^\prime)= &-& (v^\prime\cdot q)
\biggl(g^{\sigma\lambda}(p_K\cdot p_\pi) -p_\pi^\sigma
p_K^\lambda\biggr)\ -\ (q\cdot
p_\pi)\biggl(v^{\prime\sigma}p_K^\lambda-g^{\sigma\lambda}(v^\prime\cdot
p_K)\biggr)\cr &&\cr &-& q^\lambda\biggl(p_\pi^\sigma (p_K\cdot
v^\prime) - v^{\sigma} (p_K\cdot p_\pi)\biggr)\ .
\end{eqnarray}
We have not
written down other amplitudes with no contribution to the
discontinuity. Our numerical results obtained for the central
value in Eq. (\ref{fpi}) are reported in table 1.
\begin{table}[h]
  \caption{{\small Numerical values for  the imaginary part of $A_{LD}$. Units
 are GeV. First column refers to the $D, D_s$  intermediate state,
 the second column to the $D^*, D^*_s$ intermediate state.}}
\begin{center}
\begin{tabular}{|c|c|c|}
\hline $ D,\, D_s~ $  & $ D^*,\, D^*_s$ & Total
\\ \hline
$~ 1.45 \times 10^{-8}  ~$  & $ 0.89 \times 10^{-8} $ &~$ 2.34 \times 10^{-8}$ ~  \\
\hline
\end{tabular}
\end{center}
\end{table}

\section{Real part of the long-distance contribution}
\par
\noindent The real part of the diagrams \ref{ff2}c and \ref{ff2}d
could be computed by a dispersion relation, from their imaginary
parts; however this procedure suffers from the uncertainty related
to possible subtractions. A way to include them is to follow a
Feynman diagram approach, using the Effective Lagrangean discussed
in section 2. This basically amounts to including the real parts
of figs. \ref{ff2}a and \ref{ff2}b and also the other diagrams of
fig. \ref{ff2} which, as we have seen, do not contribute to the
imaginary part.

To compute $Re\ A_{LD}$ we first observe that we can change the
integration variable in (\ref{ald}) from $q=p_B - p_{D^{(*)}}$ to
the  momentum $\ell$ defined by the formula
\begin{equation}
q=p_B - p_{D^{(*)}}\equiv (m_B-m_{D^{(*)}}) v -\ell\ .
\end{equation}
We note that, by this definition, $\ell$ measures the virtuality
of the intermediate state while the velocities of the two hadrons
coincide; one can always make this choice using the
reparametrization invariance of the Heavy Quark Effective Theory
\cite{casalbuoni}. The cut-off $\mu_\ell$ corresponding to the
momentum $\ell$ can be evaluated, within our model, as follows.
For an on-shell meson with momentum $p_D=m_D\,v$, the two
constituent quarks have momenta $\vec p_c=\vec k$,  $\vec
p_q=-\vec k$ in the meson rest frame. Adding  $\ell^\mu$ shifts
$\vec p_c=\vec k\to\vec k +\vec\ell$; therefore the argument of
the wave function appearing in the calculation of the Isgur-Wise
function, instead of $|\vec k|$, is $|\vec k+\vec\ell/2|$. This
corresponds to introducing in the amplitude the form factor
$F(|\vec\ell|)$, with $F$ given in fig. \ref{ff}. Moreover, we
have to implement the condition that the residual momentum $\vec k
+\vec\ell$ of the heavy quark does not exceed the chiral symmetry
breaking scale, i.e. a mass scale around $1$ GeV. Since we assume
$<k>=350$ MeV, this gives the condition $|\vec\ell|\leq 0.65$ GeV.
The smooth form factor $F(|\vec\ell|)~( |\vec\ell|\leq 0.65~{\rm
GeV})$ can be substituted by a sharp form factor $\theta(\mu_\ell
- |\vec\ell|)$, and $\mu_\ell $ can be fixed by imposing  $\dd{
\int_0^{0.65~{\rm GeV}} F(x) dx=\int_0^{\mu_\ell} d x}$. This
procedure gives
\begin{equation}
\mu_\ell\approx 0.6~{\rm GeV}~.
\label{32}
\end{equation}
On the other hand $\ell_0$ is of the order ${\vec
\ell}^{\,2}/(2m_c)$ and is therefore negligible in the large heavy
quark mass limit. Therefore we conclude that a cut-off $\mu_\ell$
on $|\vec{\ell}|$ as given in (\ref{32}) reflects in a cut-off
$\mu^2_\ell$ on $\ell^2$. In passing, we observe that higher
derivative corrections to the Lagrangean ${\cal L}_{HH\pi}+{\cal
L}_{HH\pi\pi}$ and the weak current $L^\mu$ produce negligible
effects due to this cut-off procedure.

Having fixed the cut-off we now compute the $\ell$ integration by
performing a Wick rotation: $\ell^0\to i \ell^0$  and changing
integration variable from $|\vec\ell |$ to $L^2~=~-( (i \ell^0)^2
- \vec \ell^2 )$. We get therefore a Cottingham formula
\cite{Cottingham} as follows:
\begin{equation}
{\rm Re}\ A_{LD}=\frac{i}{2\ (2\pi)^3}
\frac{G_{F}}{\sqrt 2}V_{cb}^* V_{cs}\ a_2
\int_0^{\mu_\ell^2} dL^2
\int_{-\sqrt{L^2}}^{+\sqrt{L^2}} dl_0
\sqrt{L^2-l_0^2}\int_{-1}^{1}d \cos(\theta)\ i
\left\{\frac{j_D^\mu\ h_{D\, \mu} }{p_D^2 - m_D^2} +
\frac{\sum_{pol}\ j_{D^*}^\mu\ h_{{D^*}\, \mu} }{p_{D^*}^2 -
m_{D^*}^2} \right\}.
\end{equation}
Here $\dd j_{D^{(*)}}^\mu = <D^{(*)}|
\bar b \gamma ^\mu (1-\gamma_5) c|B>$, $\dd h_{D^{(*)}}^\mu = <K
\pi| \bar c \gamma ^\mu (1-\gamma_5) s|D^{(*)}>$. Both
$j_{D^{(*)}}^\mu$ and  $h_{D}^\mu$ can be found in
\cite{lee}\footnote{As
we have already said, we correct the heavy meson propagator to
include the hard pion momenta.}. The various contributions to
the $h_{D^*}^\mu$, corresponding to the different graphs in
fig. \ref{ff2}, are as follows:
\begin{eqnarray}
\label{e:A1}
h_{D^*\, \mu}^{(\ref{ff2}a)} & = & -i\
\frac{f_{D^*} m_{D^*}}{2\ f_\pi\ f_K} \varepsilon_\mu \; , \\
h_{D^*\, \mu}^{(\ref{ff2}b)} & = & i\ \frac{2\ \widetilde{g}
f_{D^*} m_{D^*}}{f_\pi\ f_K} \frac{(\varepsilon\cdot
p_{\pi})}{(p_{D^*}- p_\pi)^2-m_D^2}(p_{D^*}-p_{\pi})_\mu +
\nonumber \\ &&\frac{2\ \widetilde{g} f_{D^*} m_{D^*}}{f_\pi\ f_K}
\frac{1}{(p_{D^*}-
p_\pi)^2-m_{D^*}^2}\epsilon_{\mu\alpha\beta\gamma}
\varepsilon^{\alpha}\ p_\pi^\beta\ p_{D^*}^\gamma\, \; , \nn \\
h_{D^*\, \mu}^{(\ref{ff2}c)} & = & i\ \frac{4\ \widetilde{g}^2\
f_{D^*} m_{D^*}^3}{\ f_\pi\ f_K}\ \frac{ (\varepsilon \cdot
p_{\pi}) } {[(p_{D^*}- p_\pi)^2-m_D^2]\ [(p_{D^*}- p_\pi -
p_K)^2-m_{D^{*}_s}^2]}\nonumber\\ &&\left[(p_{K})_\mu -
\frac{(p_{D^*}\cdot p_K)-(p_\pi\cdot p_K)}{m_{D_s^*}^2}
(p_{D^*}-p_{\pi}-p_{K})_\mu \right ]\, \; , \nn \\
h_{D^*\, \mu}^{(\ref{ff2}d)} & = & i\ \frac{4\ \widetilde{g}^2\
f_{D^*} m_{D^*}}{\ f_\pi\ f_K} \frac{1}{[(p_{D^*}-
p_\pi)^2-m_{D^*}^2]\ [(p_{D^*}- p_\pi -
p_K)^2-m_{D^{*}_s}^2]}\nn\\ && \biggl\{ \biggl[ (p_D^*\cdot
p_\pi)(\varepsilon\cdot p_K)+ (p_\pi\cdot p_K)(\varepsilon\cdot
p_\pi) \biggr]\ (p_{D^*})_\mu +\nn\\ &&\biggl[(p_{D^*}\cdot
p_\pi-p_{D^*}^2)(\varepsilon\cdot p_K)- (p_{D^*}\cdot p_K)
(\varepsilon\cdot p_\pi)\biggr]\  (p_{\pi})_\mu+\nn\\ &&
\biggl[(p_{D^*}^2 - p_{D^*}\cdot p_\pi)(p_\pi\cdot p_K) -
(p_{D^*}\cdot p_K)(p_{D^*}\cdot p_\pi) \biggr]\ \varepsilon_\mu
\biggr\}\, \; , \nn \\
h_{D^*\, \mu}^{(\ref{ff2}e)} &=&  i \frac{ f_{D^*} m_{D^*}}{\
f_\pi\ f_K}\ \frac{p_{D^*}\cdot ( p_\pi - p_K ) } {(p_{D^*}- p_\pi
- p_K)^2-m_{D^{*}_s}^2} \left[\varepsilon_\mu
+\frac{\varepsilon\cdot(p_\pi+p_K)}{m_D^2}(p-p_\pi-p_K)_\mu\right]\,
\; , \nn \\
h_{D^*\, \mu}^{(\ref{ff2}f)} & = & \frac{4\ \widetilde{g}^2
f_{D^*} m_{D^*}}{\ f_\pi\ f_K}\ \frac{
\epsilon_{\alpha\beta\gamma\delta}\ p_{\pi}^\alpha\ p_{D^*}^\beta\
p_K^\gamma\ \varepsilon^\delta} {[(p_{D^*}- p_\pi)^2-m_{D^*}^2]\
[(p_{D^*}- p_\pi - p_K)^2-m_{D^{*}_s}^2]} \nn\\
&&(p_{D^*}-p_{\pi}-p_{K} )_\mu \; \; , \label{e:A5}
\end{eqnarray}
where $\widetilde{g} = g F(|\vec p_\pi|)$. The results for
$\mu_\ell\ =\ 0.5\div 0.7$ GeV are reported in table \ref{t:Cott}.

All the terms in the previous equations containing the factor $g$
are corrected by the form factor $F(|\vec p_\pi|)$; the terms
$h^{(2a)}$ and $h^{(2e)}$ should also contain their own
multiplicative form factors $F_a$ and $F_e$ (this holds both for
$D$ and $D^*$ intermediate states). We have not written down them
explicitly for the following reasons. $F_e$ would multiply a term
which vanishes in the chiral limit and does not affect the final
result. On the other hand $F_a(q^2)$ would represent a correction
for hard pion and kaon momenta. In the region of high $q^2$, i.e.
$q^2\in [m_D^2,(m_B-m_D)^2]$, $F_a(q^2)$ is smooth and can be put
equal to 1, i.e. to the value corresponding to the soft pion and
kaon limit. This result can be inferred  from the analogous
behaviour of the form factor $F_0^{B\to\pi}(q^2)$ describing the
coupling of a heavy and one light meson. In the high $q^2$ region
the contribution of the low lying pole $B(0^+)$ to this form
factor  vanishes in the chiral limit and  the form factor is
dominated by the direct coupling displayed in Fig. 2a, giving, as
a result, the simple formula $f_B/f_\pi$. Several numerical
analyses confirm a smooth behaviour of $F_0^{B\to\pi}(q^2)$. For
example in the quark model of \cite{polosa} $F_0^{B\to\pi}(q^2)$
increases by $40\%$ in the range $15-26$ GeV$^2$; this behaviour
can be fitted by a formula
\begin{equation}
F_0^{B\to\pi}(q^2)\propto\left(1-\frac{q^2}{(m_B+m_\Lambda)^2)}\right)^{-1}~,
\end{equation}
with $m_\Lambda\simeq 2.5$ GeV. In other models a smoother or
similar behaviour is found, see the discussion in \cite{polosa}.

For $D$ decay this formula would hold with $m_B\to m_D$:
\begin{equation}
F_0^{D\to\pi}(q^2)\propto\left (1-\frac{q^2}{(m_D+m_\Lambda)^2)}\right)^{-1}~.
\end{equation}
However, it would be a too strong assumption to assume that $F_a(q^2)$
is given by this formula; therefore we put $F_a=1$ in the sequel
and we shall enlarge the theoretical uncertainty by an extra
amount that we estimate $\pm 50\%$.

Let us finally observe that, while the numerical value of $F_a$
does not vary significantly in the range $q^2\in
[m_D^2,(m_B-m_D)^2]$, its value for  $q^2=(m_B-m_D)^2$ is formally
suppressed by one power of $(1/m_b)$ as compared to $F_a(m_D^2)$,
assuming a qualitative behaviour analogous to $F_0^{D\to\pi}(q^2)$.
\begin{center}
\begin{table}[ht!]
\caption{{\small Numerical values for the real part of $A_{LD}$
for $\mu_\ell\ =\ 0.5\div 0.7$ GeV. First column refers to the $D$
intermediate state, the second column to the $D^*$ one. Units are
GeV.}}
\begin{tabular}{|c|c|c|c|}
\hline
$\mu_\ell$ & $ D $  & $ D^* $ & Total  \\
\hline
$0.5$ & $ -4.66\times 10^{-9} $  & $ 1.62 \times 10^{-8} $ & $1.15\times 10^{-8}$ \\
\hline $0.6 $ & $ -7.77\times 10^{-9}  $  & $ 2.79\times 10^{-8}
$ & $2.01
\times 10^{-8}$ \\
\hline
$0.7$ &$ -1.19\times 10^{-8} $  & $ 4.40 \times 10^{-8} $ & $3.21\times 10^{-8}$\\
\hline
\end{tabular}
\label{t:Cott}
\end{table}
\end{center}

\section{Conclusions}
Our numerical results show that the long-distance charming penguin
contributions to the decays  $B \to K \pi$ are significant. These
results agree qualitatively  with a phenomenological analysis of
these contributions  given in \cite{Ciuchini1}. In particular, we
found that the absorptive part due to the $D,D_s$ states is
somewhat bigger than that from the  $D^{*},D_s^{*}$ states, but of
the same sign. The real part due to the $D^{*},D_s^{*}$
states is however $3-4$ times bigger and opposite in sign to the
contributions from the $D,D_s$ states. As shown in table 1 and
table 2, the total contribution for the real part and absorptive
part are of the same order of magnitude, at the $10^{-8}\rm\, GeV$
level. The results for the branching ratios are as follows:
\begin{eqnarray}
{\cal B}(B^+ \to K^0\pi^+)&=&(2.4^{+2.7}_{-1.9})\times10^{-5} \cr
{\cal B}(B^0 \to K^+\pi^-)&=&(1.5^{+1.8}_{-1.3})\times10^{-5}~.
\label{risultati}
\end{eqnarray}
The central values in this equation correspond to the central
values of the parameters $g$, $<k>$ (i.e. $F{|\vec p_\pi|}$) and
$\mu_\ell$. The theoretical uncertainties in the branching ratios
are obtained by varying the parameters in the ranges $g=0.40\pm
0.08$, $<k>=(350\pm 50)$ GeV and $\mu_\ell=(0.6\pm 0.1)$ GeV. An
extra theoretical uncertainty of $50\%$ has been added in
quadrature to $Re A_{LD}$. The results in (\ref{risultati}) agree with the
experimental values in eq. (\ref{CLEO2}).

The existence of two different contributions with different strong
and weak phases produces a direct CP violation in the decay
$B\to K^\pm\pi^\mp$. As a matter of fact, we find for
\begin{equation}
{\cal A}=\frac{\Gamma(\bar B^0\to K^-\pi^+)-\Gamma(B^0\to
K^+\pi^-)}{\Gamma(\bar B^0\to K^-\pi^+)+\Gamma(B^0\to K^+\pi^-)}
\end{equation}
the value ${\cal A}=+0.21$ (for $\gamma=54.8^0$), which is
compatible with the recent results from CLEO \cite{cleocp}.
On the other hand the present model produces no CP asymmetries for
charged $B$ decays into $K\pi$.

Let us finally discuss the problem of the scaling of our results
with $m_b$. Assuming, as in \cite{Beneke}, that $F_0^{B\to\pi}(0)$
scales as $\dd\left(\frac{\Lambda_{QCD}}{m_b}\right)^{3/2}$,
$A_{SD}$ scales as $m_b^{1/2}$. On the other hand the charming
penguin contribution $A_{LD}$ is non leading due to the various
form factors which vanish for $m_b,\,m_c\to\infty$. Therefore the
following conclusion can be drawn. The  charming penguin
contributions violate naive factorization and are of leading order
in $\alpha_s$, as they arise from non perturbative calculations;
nevertheless they do not contradict the results of
\cite{Beneke,keum} since they are suppressed in the
infinite heavy quark mass limits. In spite of that, as $m_b,m_c$
in reality are finite, the charming penguin numerical contribution
to the branching ratios in $B\to K\pi$ decays is significant,
basically due to the enhancement of the CKM matrix elements.

In conclusion, we believe that the charmed resonance contributions
we found seem to be capable of producing the charming-penguin
terms suggested in \cite{Ciuchini1}  within theoretical errors.
This method could then be applied also to other two-body
nonleptonic $B$ decay channels, such as the $B \to \pi \pi$ and
the $B \to K \eta ^\prime$. We only mention here a result for the
charming-penguin contribution in ${B}^{0} \to \pi^{+}\pi^{-} $
decay. This contribution can be obtained from the results we
obtained for ${B}^{0} \to K^{+}\pi^{-} $ using  $SU(3)$ symmetry
for the weak current matrix elements
$<K^{+}\pi^{-}|(\bar{c}s)_{L}|D_{s}>$  and
$<\pi^{-}\pi^{+}|(\bar{c}d)_{L}|D^{-}>$. Thus, in the $SU(3)$
limit(by ignoring the $K-\pi$ mass difference) \cite{lee}
\begin{equation}
A_{\rm LD}({B}^{0} \to \pi^{+}\pi^{-}) =
(V_{cd}/V_{cs})(f_K/f_\pi) \times A_{\rm LD}({B}^{0} \to
K^{+}\pi^{-}) ~.
\label{Bpipi}
\end{equation}
Since the ${B}^{0} \to
\pi^{+}\pi^{-} $ decay is dominated by the tree-level operators,
this charming-penguin contribution behaves like the penguin terms
and reduces the ${B}^{0} \to \pi^{+}\pi^{-} $ decay rate by a
small amount. The details of this work will be given elsewhere
together with the results for $B \to K \eta ^\prime$.

{\it Acknowledgements} We thank F. Ferroni for useful correspondence
on the BaBar data.

\end{document}